\documentclass[12pt]{article}

\usepackage{amssymb,amsmath,graphics}

\numberwithin{equation}{section}

\allowdisplaybreaks

\newenvironment{Proof}{\removelastskip\par\medskip
\noindent{\em Proof.}
\rm}{\penalty-20\null\hfill$\square$\par\medbreak}

\def\real{{\mathord{\mathbb R}}}

\newtheorem{prop}{Proposition}
\newtheorem{lemma}{Lemma}

\newtheorem{remark}{Remark}
\def\Dom{{\mathrm{{\rm Dom ~}}}}

\begin{document}
 ~
\begin{center}
{\Large On the pricing and hedging of options for highly volatile periods} \\
\bigskip
{\large Youssef El-Khatib}\footnote{UAE University, Department of Mathematical Sciences, Al-Ain, P.O. Box 17551. United Arab Emirates.{\ E-mail : Youssef\_Elkhatib@uaeu.ac.ae.}
}\ \ \ \ \ {\large Abdulnasser Hatemi-J}\footnote{UAE University, Department of Economics and Finance, Al-Ain, P.O. Box 17555, United Arab Emirates. {\ E-mail : Ahatemi@uaeu.ac.ae.}}
\end{center}
~
\begin{abstract}
Option pricing is an integral part of modern financial risk management. The well-known Black and Scholes (1973) formula is commonly used for this purpose. This paper is an attempt to extend their work to a situation in which the unconditional volatility of the original asset is increasing during a certain period of time. We consider a market suffering from a financial crisis. We provide the solution for the equation of the underlying asset price as well as finding the hedging strategy. In addition, a closed formula of the pricing problem is proved for a particular case. The suggested formulas are expected to make the valuation of options and the underlying hedging strategies during financial crisis more precise.
\end{abstract}

\baselineskip=0.5cm

\baselineskip=0.5cm \noindent\textbf{Keywords:} Options Pricing and Hedging, Financial Crisis, Black and Scholes Formula.
\baselineskip0.7cm
\section{Introduction}
In order to neutralize or at least try to reduce the price risk of
financial assets such as stocks, financial derivatives such as
options are regularly utilized. The well-known Black and
Scholes (1973) option pricing formula, denoted as BS henceforth, is
routinely used for this purpose. The aim of current paper attempts to extend the work of BS in order to account for the possibility that the unconditional volatility of the underlying asset increases during a certain period before maturity. It is a well-established fact that the volatility of financial assets tend to increase during a financial crisis period. Due to the increasingly dominant globalization effect of the financial markets, the likelihood of spillover effects and the resulting contagion is higher than ever. As a consequence, the BS formula might not perform accurately during a financial crisis. A particular event that can support the view that the BS formula performs well when the market is doing well but not during a financial crisis is the performance of a well-known hedge fund entitled Long-Term Capital Management (LTCM). This hedge fund was established by Scholes and Merton in 1994. The LTCM performed extremely well and provided returns over 40\% per year until the 2007 East Asian financial crisis combined with the Russian financial crisis in 2008 resulted in a loss of $4.6$ billion dollars within four months. This event caused the LTCM to go bankrupt. \newline

The current paper provides a solution for the equation of the underlying asset price in a market with increasing unconditional volatility of the asset across time as well as finding the hedging strategy. In addition, a closed formula of the pricing problem is proved for a particular case. The existing option pricing models originating from empirical studies on the dynamics of financial markets after the occurrence of a financial crash do not seem to accord with the stochastic models. For instance, while the BS model \cite{BlackScholes} assumes
that the underlying asset price follows a geometric Brownian motion, however, the work of \cite{sornette2003} shows empirically that the post-crash dynamics follow a converging oscillatory motion. In addition, the paper of \cite{lillo2003} shows that financial
markets follow power-law relaxation decay. Several ideas have been
suggested to overcome this shortcoming of the BS model.
In fact, new option pricing models have been developed based on
empirical observations (see for instance \cite{savit1989},
\cite{deeba2002}, \cite{tvedt1998}, \cite{dibeh2005} and
\cite{mccauley2004}). Recently, \cite{dibeh2007} suggests a newer model which is claimed to extend the BS model. The
extension attempts to take into account the post-crash dynamics as proposed by
\cite{sornette2003}. The authors utilize the following stochastic
differential equation that couples the post-crash market index to
individual stock price $(S_t)_{0\leq t \leq T}$ via the function $g(t)$
$$\frac{dS_{t}}{S_{t}}=\left(a+\frac{bg(t)}{S_t}\right)dt+\left(\sigma + \frac{\beta g(t)}{S_t}\right)dW_{t},$$
where $t\in \lbrack 0,T]$, $S_{0}=x>0$ and $g(t)=A+Be^{\alpha t}sin(\omega t)$.
The values $a$, $b$, $\beta$, $A$ and $B$ are real constants. The volatility of the original asset is denoted by $\sigma$.
The authors obtain the following partial differential equation (P.D.E.) for the option price
$$
\frac{\partial C}{\partial t}+r S\frac{\partial C}{\partial S}-r C+\frac{1}{2}\left(\sigma S + \beta g(t)\right)^{2}\frac{\partial^2 C}{\partial S^2}=0,
$$
with the terminal condition $C(S,T) =(S-K)^{+}$. Where $C$ is the
call option's price, $r$ is the risk free rate, and $K$ is the
strike price. \newline
The authors do not however provide any solution for the suggested model. A solution that is provided in this paper is utilized to derive an alternative option pricing formula. For another recent approach on options pricing see \cite{GLZ2012}.\\

The remaining part of the paper is structured
as follows. In Section 2 we present the model and suggest a solution for the model combines with the proof. Section 3 derives
and mathematically proves an alternative formula for pricing and hedging options. The last section concludes the paper.
\section{The Model}
In order to derive the option pricing formula we need to make the
following assumptions, in line with BS:
\begin{enumerate}
  \item The short-run risk free rate, $r$, is known and it is assumed to be constant.
  \item The distribution of stock prices within any finite interval is assumed to be lognormal.
  \item No dividends are paid out during the life time of the option.
  \item No transaction costs prevail.
  \item Short selling opportunities exist.
\end{enumerate}
However, unlike BS the variance of the original
asset does not need to be constant but increasing across the time
span. Assume that the probability space is $(\Omega,{\cal F},P)$. Assume also that $(W_t)_{t \in [0,T]}$ is a Brownian motion process and $({\cal F}_t)_{t \in [0,T]}$ be the natural filtration generated by $(W_t)_{t\in [0,T]}$.
We consider a market with two assets: a risky asset $S=(S_t)_{t\in[0,T]}$ to which is related an European call option and a riskless one given by
$$ dA_{t}=r A_{t}dt,\ \ \ t\in \lbrack 0,T],\ \ \ A_{0}=1.
$$
Assume that $P$ is the risk-neutral probability and that, under $P$, the data generating process for the stock price at
time $t$, denoted by $S_t$, is the following stochastic differential equation that accounts of the post-crash (crisis) effect
\begin{equation}
\label{eq10}
dS_{t}= r S_t dt+(\sigma S_t + \beta A(t))dW_{t},
\end{equation}
where $t \in [0,T]$, $S_0=x>0$ and $\beta$ is a constant. The denotation $\sigma$ signifies the volatility of the original asset.\newline
The previous model is a special case of the model considered in \cite{dibeh2007}
\begin{equation}
\label{eq11}
dS_{t}= r S_t dt+(\sigma S_t + \beta g(t))dW_{t},
\end{equation}
where $g(t)$ is a deterministic function\footnote{For instance, one can take $g(t)=A+Be^{\alpha t}sin(\omega t)$ as in \cite{dibeh2007}.}.
Recall that a stochastic process is a function of
two variables i.e. time $t \in [0,T]$ and the event $\omega \in
\Omega$. However, in the literature it is common to write $S_t$ instead of
$S_t(\omega)$. The same is true for $W_t$ or any other stochastic process mentioned in this paper.\newline
Let $(\xi_t)_{t \in [0,T]}$ be the stochastic process defined by
\begin{equation}
\label{eq_xit}
d\xi_{t}= r \xi_t dt+\sigma \xi_t dW_t,\ \ \ \xi_0=1.
\end{equation}
The solution of the equation (\ref{eq11}) is given by the following proposition
\begin{prop}
\label{prop_sol}
For $0\leq t\leq T$, we let $\xi_{t}:=\exp\left[\left(r-\frac{\sigma^2}{2}\right)t+\sigma W_t\right].$ The solution of equation~(\ref{eq11}) is given by
\begin{equation}
\label{St}
S_t=x\xi_{t}-\frac{\beta}{\sigma}\left(g(t)+\int_0^t \xi_t {\xi_s}^{-1}(r g(s)-g^{'}(s))ds\right).
\end{equation}
\end{prop}
\begin{Proof}
It is well-known that $\xi_{t}$ as a geometric Brownian motion satisfies the following stochastic differential equation
$$d\xi_{t}= r\xi_{t}dt+\sigma \xi_{t}dW_t, \   \  \  \xi_0=1.$$ Moreover, by applying the It\^o's formula to ${\xi_t}^{-1}:=\exp\left[\left(-r+\frac{\sigma^2}{2}\right)t-\sigma W_t\right]$, we obtain
\begin{eqnarray}
\nonumber
d({\xi_t}^{-1})&=& \frac{-1}{{\xi_t}^{2}}d\xi_{t}+ \frac{1}{{\xi_t}^{3}}(\sigma \xi_{t})^2 dt= \frac{-1}{{\xi_t}^{2}}(r\xi_{t}dt+\sigma \xi_{t}dW_t)+ \frac{\sigma^2}{{\xi_t}} dt\\
\label{xi-1}
&=& {\xi_t}^{-1}[(-r+\sigma^2)dt-\sigma dW_t].
\end{eqnarray}
In the particular case when $\beta=0$, the solution is  $S_t =x\xi_{t}$. However, if $\beta \neq 0$, we need to use the variation of the constants method,  so we search for a solution in the form of $S_t =\xi_{t}Y_t$, with $Y_0=S_0=x$. Thus, we have
\begin{eqnarray*}
dY_t&=&d(S_t {\xi_t}^{-1})={\xi_t}^{-1} dS_t +S_t d({\xi_t}^{-1})+d[S_t, {\xi_t}^{-1}]\\
&=&{\xi_t}^{-1}(r S_t dt+(\sigma S_t + \beta g(t))dW_{t})+S_t{\xi_t}^{-1}((-r+\sigma^2)dt-\sigma dW_t)\\
&&+ \left[(\sigma S_t + \beta g(t))dW_{t},-{\xi_t}^{-1}\sigma dW_t\right]\\
&=&-\sigma \beta g(t){\xi_t}^{-1}dt+\beta g(t){\xi_t}^{-1}dW_{t}\\
&=&-\frac{\beta}{\sigma}g(t)\left(\sigma^2 {\xi_t}^{-1}dt-\sigma{\xi_t}^{-1}dW_{t}\right).
\end{eqnarray*}
Using equation (\ref{xi-1}) and the integration by parts, we have
\begin{eqnarray*}
dY_t&=&-\frac{\beta}{\sigma}g(t)\left(d({\xi_t}^{-1})+r{\xi_t}^{-1}dt\right)\\
&=&-\frac{\beta}{\sigma}\left(g(t)d({\xi_t}^{-1})+r g(t){\xi_t}^{-1}dt\right)\\
&=&-\frac{\beta}{\sigma}\left(d(g(t){\xi_t}^{-1})-{\xi_t}^{-1}d g(t)+r g(t){\xi_t}^{-1}dt\right)\\
&=&-\frac{\beta}{\sigma}\left(d(g(t){\xi_t}^{-1})-{\xi_t}^{-1}g^{'}(t)d t+r g(t){\xi_t}^{-1}dt\right)\\
&=&-\frac{\beta}{\sigma}\left(d(g(t){\xi_t}^{-1})-{\xi_t}^{-1}(r g(t)-g^{'}(t))dt\right)\\
\end{eqnarray*}
Therefore,
\begin{equation}
\label{Y_t}
Y_t=x-\frac{\beta}{\sigma}\left(g(t){\xi_t}^{-1}+\int_0^t{\xi_s}^{-1}(r g(s)-g^{'}(s))ds\right).
\end{equation}
The solution of $S$ is then given by
\begin{eqnarray*}
 S_t &=&\xi_{t}Y_t=\xi_{t}\left(x-\frac{\beta}{\sigma}\left(g(t){\xi_t}^{-1}+\int_0^t{\xi_s}^{-1}(r g(s)-g^{'}(s))ds\right)\right)\\
&=&x\xi_{t}-\frac{\beta}{\sigma}\left(g(t)+\int_0^t \xi_t {\xi_s}^{-1}(r g(s)-g^{'}(s))ds\right).
\end{eqnarray*}
\end{Proof}
The previous equation could take negative values which is not suitable for stock price values. In order to overcome this shortcoming in our model for the asset, we need the following lemma.
\begin{lemma}
\label{lemma_1}
The stock price modeled by equation (\ref{eq10}) is bounded as follows
\begin{equation}
xe^{(r-\frac{\sigma^2}{2})t-3\sigma \sqrt{t} }- \frac{\beta}{\sigma}e^{rt} \leq  S_{t}  \leq xe^{(r-\frac{\sigma^2}{2})t+3\sigma \sqrt{t} }- \frac{\beta}{\sigma}e^{rt},
\end{equation}
with the probability of $99.6\%$.
\end{lemma}
\begin{Proof}
It is well-known that if $X$ is a random variable that follows a normal distribution, i.e. $X \sim N(\mu, \sigma^2)$, then $\mu-3\sigma \leq X \leq \mu+3\sigma$ with the probability $99.6\%$. Since $\frac{W_t}{\sqrt{t}}$ follows $N(0,1)$, then we have
\begin{eqnarray*}
(r-\frac{\sigma^2}{2})t-3\sigma \sqrt{t} \leq & (r-\frac{\sigma^2}{2})t+\sigma W_t& \leq (r-\frac{\sigma^2}{2})t+3\sigma \sqrt{t}\\
e^{(r-\frac{\sigma^2}{2})t-3\sigma \sqrt{t}}\leq & \xi_{t} & \leq e^{(r-\frac{\sigma^2}{2})t+3\sigma \sqrt{t}}\\
xe^{(r-\frac{\sigma^2}{2})t-3\sigma \sqrt{t} }- \frac{\beta}{\sigma}e^{rt} \leq & S_{t} & \leq xe^{(r-\frac{\sigma^2}{2})t+3\sigma \sqrt{t} }- \frac{\beta}{\sigma}e^{rt},
\end{eqnarray*}
with $99.6\%$ probability.
\end{Proof}
Now we can state the following proposition:
\begin{prop}
\label{prop_beta}
If for $0\leq t\leq T$, we choose $\beta$ such that
$$
 \beta \leq x\sigma e^{-(\frac{\sigma^2}{2}T+3\sigma \sqrt{T})}
$$
then $S_t>0$ with $99.6\%$ probability.
\end{prop}
\begin{Proof} It is clear from lemma~\ref{lemma_1} that if $xe^{(r-\frac{\sigma^2}{2})t-3\sigma \sqrt{t} }- \frac{\beta}{\sigma}e^{rt}>0$ then $S_t>0$ with probability $99.6\%$. But $xe^{(r-\frac{\sigma^2}{2})t-3\sigma \sqrt{t} }- \frac{\beta}{\sigma}e^{rt}>0$ is equivalent to $\beta<x\sigma e^{-(\frac{\sigma^2}{2}t+3\sigma \sqrt{t})}$. The function $f(t):=e^{-(\frac{\sigma^2}{2}t+3\sigma \sqrt{t})}$ is decreasing and its minimum on the interval $[0,T]$ is $e^{-(\frac{\sigma^2}{2}T+3\sigma \sqrt{T})}$. This ends the proof.
\end{Proof}
 Hence for being almost sure (with $99.6\%$ probability) that the stock price model is positive, we impose to $\beta$ the condition in proposition~\ref{prop_beta}. If, under this condition, at a certain time between $0$ and $T$ the financial asset obtains a negative value, we assume in this case that the company emitting the underlying asset goes bankrupt and thus the asset and its related options are no more tradeable on the financial market.
\section{The Option Valuation and Hedging Formulae}
In this section, we provide the hedging formula for the European options with underlying asset represented by equation (\ref{St}). Then we assume $g(t)=A_t$, which allows us to provide a Black Scholes-like pricing formula.
\subsection{The hedging strategy}
We are interested in finding the hedging strategy
for our model (\ref{St}). Let $\eta_t$ and $\zeta_t$ denote the number of units
  invested at time $t$
  in the risky and risk-less assets respectively. Thus, the value
 $V_t$ of the portfolio at time $t$ is given by
  \begin{equation}
 \label{e43}
 V_t = \zeta_{t } A_t + \eta_t S_t, \ \ \ t\in [0,T].
 \end{equation}
We assume that the portfolio is self-financing, i.e.
$$dV_t = \zeta_t dA_t  + \eta_t dS_t,\ \ \ t\in [0,T],$$
therefore,
\begin{equation}
 \label{e330}
 dV_t = r V_t dt + \eta_t (\sigma S_t +\beta g(t)) dW_t, \ \ \ t\in
[0,T].
\end{equation}
and
\begin{equation}
 \label{e440}
 V_T e^{-rT} = V_{0} + \int_0^T
  \eta_t (\sigma S_t +\beta g(t)) e^{-rt} dW_t.
\end{equation}
We seek a portfolio $(\zeta_t,\eta_t)_{t\in [0,T]}$
which leads to the payoff $V_T = h(S_T)$, for instance for call options $h(S_T)=(S_T -K)^{+}$ and for put options $h(S_T)=(K-S_T)^{+}$. We assume that $V_t=C(S_t,t)$. The following proposition gives the replicating portfolio for European options.
\begin{prop}
\label{lbk}
 The replicating portfolio of an European call option
 is given by
 \begin{equation}
\label{lkh}
\eta_t= C_x(S_t,t)= e^{-r(T-t)}E[\xi_{T-t} 1_{[K , \infty [}(S_T)\mid{\cal F}_t ], \ \ \ t \in [0,T],
\end{equation}
\end{prop}
\begin{Proof}
Applying It\^o formula to $V_T=C(T,S_T)$ and using equation (\ref{eq11}) we obtain
\begin{eqnarray*}
dC(S_t,t)&=&\partial_t C(S_t,t)dt+\partial_x C(S_t,t)dS_t
+\frac{1}{2}\partial^2_{xx} C(S_t,t)d\langle S_t,S_t\rangle\\
&=&C_t(S_t,t)dt+ C_x (S_t,t)\left(r S_t dt+(\sigma S_t + \beta g(t))dW_{t}\right)\\
&&+\frac{1}{2}C_{xx}(S_t,t)(\sigma S_t + \beta g(t))^2 dt\\
&=&\left[C_t(S_t,t)+r S_t C_x (S_t,t) +\frac{1}{2}C_{xx}(S_t,t)(\sigma S_t + \beta g(t))^2\right] dt\\
&&+( \sigma S_t + \beta g(t)) C_x(S_t,t)dW_t,
\end{eqnarray*}
then the last equation compared with equation (\ref{e330}) gives
$$
\eta_t (\sigma S_t +\beta g(t))=(\sigma S_t + \beta g(t)) C_x(S_t,t),
$$
which implies the first part of the equality (\ref{lkh}). The second part is obtained
by the Clark-Ocone formula. Let $D_t$ denote the Malliavin derivative\footnote{for more details on Malliavin derivative, we refer the reader for instance to \cite{oksendal1996}.} on the Wiener space. as follows
$$
(S_T -K)^{+}
 = E\left[ (S_T -K)^{+} \right]
 + \int_0^T E\left[D_t (S_T -K)^{+} \mid
 {\cal F}_t \right]dW_t,$$
and comparing with (\ref{e440}) we obtain
 \begin{eqnarray}
 \nonumber
  V_{0} & = & e^{-rT} E[F], \\
\label{hed}
 \eta_t & = & (\sigma S_t +\beta g(t))^{-1} E[D_t (S_T -K)^{+} \mid{\cal F}_t ]
  e^{-r(T-t)}, \ \ \ t\in [0,T].
 \end{eqnarray}
 We have
 \begin{eqnarray}
\nonumber
 D_t S_T  &= & D_t\left[x\xi_{T}-\frac{\beta}{\sigma}\left(g(T)+\xi_T\int_0^T {\xi_s}^{-1}(r g(s)-g^{'}(s))ds\right)\right]\\
 \nonumber
 &=&x\sigma \xi_T -\frac{\beta}{\sigma}\left(\sigma \xi_T \int_0^T {\xi_s}^{-1}(r g(s)-g^{'}(s))ds -\sigma \xi_T \int_t^T {\xi_s}^{-1}(r g(s)-g^{'}(s))ds\right)\\
  \label{e47}
 &=&(\sigma S_{t} + \beta g(t))\xi_{T-t}, \ \ \ t\in [0,T].
\end{eqnarray}
 The chain rule $Df(F) = f'(F)DF$
 holds for $F\in {\cal S}$ and $f\in {\cal C}^2_b (\real)$.
 We may approach $x\mapsto (x-K)^+$ by polynomials
 on compact intervals and proceed e.g. as in \cite{oksendal1996}.
 By dominated convergence,
 $(S_T - K)^+\in \Dom (D)$ and (\ref{e47}) can be written as
$$
 D_t  (S_T - K)^+
 = (\sigma S_{t} + \beta g(t))\xi_{T-t} 1_{[K , \infty [}(S_T),\ \ \ 0\leq t \leq T.
$$
Then,
\begin{eqnarray*}
\eta_t & = & (\sigma S_t +\beta g(t))^{-1} E[(\sigma S_{t} + \beta g(t))\xi_{T-t} 1_{[K , \infty [}(S_T)\mid{\cal F}_t ]\\
&=& E[\xi_{T-t} 1_{[K , \infty [}(S_T)\mid{\cal F}_t ].
\end{eqnarray*}
The proof is completed.
\end{Proof}
\begin{remark}
Note that the replication portfolio of an European put option can be obtained by the same arguments of the previous proposition as follows
\begin{equation}
\label{lkh2}
\eta_t= P_x(S_t,t)= e^{-r(T-t)}E[\xi_{T-t} 1_{[0 , K]}(S_T)\mid{\cal F}_t ], \ \ \ t \in [0,T],
\end{equation}
where $P_x(S_t,t)$ is the price of European put.
\end{remark}
\subsection{The Option Valuation Formula when $g(t)=A_t=e^{rt}$}
We assume that $g(t)=A_t=e^{rt}$, which results in $r g(t)-g^{'}(t)=0$. And thus the dynamic of the price process in equation (\ref{St}) becomes the following:
\begin{equation}
\label{St2}
S_t=x\xi_{t}-\frac{\beta}{\sigma}g(t).
\end{equation}
The next proposition gives the premium (price at $t=0$) of an European call option based on our approach.
\begin{prop}
Assume that $g(t)=e^{rt}$, so that the dynamic of the price process, $S_T$, is given by (\ref{St2}), then the premium of an European call option with strike $K$ is given by
\begin{equation}
\label{premium1}
C(S_T, K)=E[e^{-rT}(S_T -K)^{+}]=S_0 \Phi (d_1^{\beta}) -\left(K+\frac{\beta}{\sigma}e^{rT}\right)e^{-rT}\Phi (d_2^{\beta}),
\end{equation}
where
\begin{equation}
\label{d1}
d_1^{\beta}= \frac{1}{\sigma \sqrt{T}}\left( \ln\left(\frac{S_0}{K+\frac{\beta}{\sigma}e^{rT}}\right)+(r+\frac{\sigma}{2})T\right),
\end{equation}
and
\begin{equation}
\label{d2}
d_2^{\beta}=  \frac{1}{\sigma \sqrt{T}}\left( \ln\left(\frac{S_0}{K+\frac{\beta }{\sigma}e^{rT}}\right)+(r-\frac{\sigma}{2})T\right),
\end{equation}
and $\Phi (d)=\int_{-\infty}^d \frac{e^{-u^{2}/2}}{\sqrt{2\pi}}du.$
\end{prop}
\begin{Proof}
Let $C(x\xi_T, K)$ be the price of an European option with underlying price $x\xi_T$ and strike price $K$. Then, the BS formula gives
\begin{equation}
\label{premium0}
C(x\xi_T, K)=E[e^{-rT}(x\xi_T -K)^{+}]=S_0 \Phi (d_1) -K e^{-rT}\Phi (d_2),
\end{equation}
where
$$
d_1=  \frac{1}{\sigma \sqrt{T}}\left( \ln\left(\frac{S_0}{K}\right)+(r+\frac{\sigma}{2})T\right)\ \ \ \mbox{and} \ \ \ d_2=  \frac{1}{\sigma \sqrt{T}}\left( \ln\left(\frac{S_0}{K}\right)+(r-\frac{\sigma}{2})T\right),
$$
and $\Phi (d)=\int_{-\infty}^d \frac{e^{-u^{2}/2}}{\sqrt{2\pi}}du.$ \newline
The price of an European option with underlying price $S_T$ (given by equation (\ref{St})) and strike price $K$ is then
\begin{eqnarray*}
C(S_T, K)&=&E[e^{-rT}(S_T -K)^{+}]=E\left[e^{-rT}\left(x\xi_{T}-\frac{\beta}{\sigma}e^{rT} -K\right)^{+}\right]\\
&=&E\left[e^{-rT}\left(x\xi_{T}-\left(\frac{\beta}{\sigma}e^{rT} +K\right)\right)^{+}\right]=E[e^{-rT}(x\xi_T -K^{'})^{+}],
\end{eqnarray*}
where $K^{'}=\frac{\beta}{\sigma}e^{rT}+K$. Now, by (\ref{premium0}), we obtain
$$C(S_T, K)=E[e^{-rT}(x\xi_T -K^{'})^{+}]=C(x\xi_T, K^{'})=S_0 \Phi (d_1^{\beta}) -\left(K+\frac{\beta}{\sigma}e^{rT}\right)e^{-rT}\Phi (d_2^{\beta}),$$
where $d_1^{\beta}$ and $d_2^{\beta}$ are given by equations (\ref{d1}) and (\ref{d2}). The proof is completed.
\end{Proof}
Here $K^{'}$, which is greater than $K$, can be seen as a new strike price. So, during crisis, the price of an European call option can be seen as the price of a new option with the same parameters but with a higher strike price. Since the price of the European call option is inversely related to the strike price, the call option price decreases during a financial crisis.\newline
Concerning the price of an European put option, one can use the Put-Call parity relation for European options:
\begin{equation}
\label{putcall}
S_0+P(S_T, K)=C(S_T, K)+Ke^{-rT}.
\end{equation}
Based on this condition, the following proposition can be used to determine the premium of an European put option.
\begin{prop}
Assume that $g(t)=e^{rt}$, so that the dynamic of the price process, $S_T$, is given by (\ref{St2}), then the premium of an European put option with strike $K$ is given by
\begin{equation}
\label{premium2}
P(S_T, K)=S_0 \Phi (d_1^{\beta}) -\left(K+\frac{\beta}{\sigma}e^{rT}\right)e^{-rT}\Phi (d_2^{\beta})+Ke^{-rT}-S_0,
\end{equation}
where $d_1^{\beta}$ and $d_2^{\beta}$ are given by equations (\ref{d1}) and (\ref{d2})
and $\Phi (d)=\int_{-\infty}^d \frac{e^{-u^{2}/2}}{\sqrt{2\pi}}du.$
\end{prop}
\begin{Proof}
The proof is straightforward by equation (\ref{putcall}) and (\ref{premium1}).
\end{Proof}
Let $(\xi_{t,u}^x )_{u\in [t,T]}$ be the process defined as
 $$
 d\xi_{t,u}^x = r \xi_{t,u}^x du  + \sigma \xi_{t,u}^x dW_u, \ \ \ u\in
 [t,T], \ \ \ \xi_{t,t}^x = x.$$
We have $\xi_{t} = \xi_{0,t}^1$, $t\in [0,T]$.
The next proposition gives the price of the European call option at any time $t$ based on our approach.
\begin{prop}
\label{lbk2}
Assume that $g(t)=e^{rt}$, so that the dynamic of the price process, $S_T$, is given by (\ref{St2}), then the price of an European call option and the price of European put option with strike $K$ at time $t\in [0,T]$ are respectively given by
$$
C(t,S_t)=S_t \Phi (d_{t,1}^{\beta}) -\left(K+\frac{\beta}{\sigma}e^{r(T-t)}\right)e^{-r(T-t)}\Phi (d_{t,2}^{\beta}),
$$
and
$$
P(t,S_t)=S_t \Phi (d_{t,1}^{\beta}) -\left(K+\frac{\beta}{\sigma}e^{r(T-t)}\right)e^{-r(T-t)}\Phi (d_{t,2}^{\beta})+Ke^{-r(T-t)}-S_t,
$$
where
\begin{equation}
\label{dt2}
d_{t,1}^{\beta}= \frac{1}{\sigma \sqrt{T-t}}\left( \ln\left(\frac{S_t}{K+\frac{\beta}{\sigma}e^{r(T-t)}}\right)+(r+\frac{\sigma}{2})(T-t)\right),
\end{equation}
and
$$
d_{t,2}^{\beta}=  \frac{1}{\sigma \sqrt{T-t}}\left( \ln\left(\frac{S_t}{K+\frac{\beta }{\sigma}e^{r(T-t)}}\right)+(r-\frac{\sigma}{2})(T-t)\right).
$$
\end{prop}
\begin{Proof}
By using the Markov property of the process $(S_t)_{t\in [0,T]}$, we have the following result
\begin{eqnarray*}
C(t,S_t)&=&e^{-r(T-t)}E\left[(S_T - K)^{+} \mid {\cal F}_t \right]\\
        &=& e^{-r(T-t)} E \left[(\xi_{t,T}^x-(\frac{\beta}{\sigma}g(T)-K))^{+}\right]_{ x = S_t}\\
        &=&C(\xi_{t,T}^x, K^{'})_{ x = S_t},
\end{eqnarray*}
where $K^{'}=\frac{\beta}{\sigma}e^{rT}+K$, which ends the proof.
Similarly, one can use equation~(\ref{premium2}) and the Markov property of the process $(S_t)_{t\in [0,T]}$ to obtain the price of the European put option. The proof is completed.
\end{Proof}
\begin{remark}
Note that the replication portfolio of an European option when $g(t)=e^{rt}$ can be completely obtained using proposition.~\ref{lbk} and proposition.~\ref{lbk2}, for instance in the case of European call options we have
$$
\eta_t= C_x(S_t,t)=\Phi (d_{t,1}^{\beta}),
$$
where $d_{t,1}^{\beta}$ is given by equation~(\ref{dt2}).
\end{remark}
\section{Conclusions}
This article reinvestigates the issue of option pricing by extending
the seminal work of Black and Scholes (1973) to cases in which the
unconditional volatility of the original assets can increase across the time span.
This scenario is expected to depict a realistic situation in which
the financial market is characterized by a crisis. The Black and Scholes
formula, which does not take into account the impact of the increase in
volatility during the crisis, is likely to not perform accurately. We
offer and mathematically prove an alternative formula for option
pricing during periods in which the market is under stress. The suggested formula
can be used for hedging purposes also. Thus, this formula is expected to make the valuation of options more
accurate especially during a financial crisis, in which the need for
more accurate evaluations is urgent.
\bibliographystyle{plain}
\bibliography{Mybib}
\end{document}